\documentclass[twocolumn]{aastex631}

\usepackage[utf8]{inputenc}
\usepackage[draft]{fixme}
\usepackage{graphicx}
\usepackage{graphicx,xspace}
\usepackage{hyperref}
\usepackage{breakurl}
\usepackage{soul}
\usepackage{amsthm}
\usepackage{amsmath}

\usepackage{graphicx}	

\shorttitle{Probing the EBL with HAWC}
\shortauthors{THE HAWC COLLABORATION}
\begin{document}

\title{Probing the extragalactic mid-infrared background with HAWC}

\correspondingauthor{M. Fernández Alonso}
\email{mkf5479@psu.edu}

\author[0000-0003-0197-5646]{A.~Albert}
\affiliation{Physics Division, Los Alamos National Laboratory, Los Alamos, NM, USA }

\author[0000-0001-8749-1647]{R.~Alfaro}
\affiliation{Instituto de F\'{i}sica, Universidad Nacional Autónoma de México, Ciudad de Mexico, Mexico }

\author{C.~Alvarez}
\affiliation{Universidad Autónoma de Chiapas, Tuxtla Gutiérrez, Chiapas, México}

\author{J.C.~Arteaga-Velázquez}
\affiliation{Universidad Michoacana de San Nicolás de Hidalgo, Morelia, Mexico }

\author[0000-0002-4020-4142]{D.~Avila Rojas}
\affiliation{Instituto de F\'{i}sica, Universidad Nacional Autónoma de México, Ciudad de Mexico, Mexico }

\author[0000-0002-2084-5049]{H.A.~Ayala Solares}
\affiliation{Department of Physics, Pennsylvania State University, University Park, PA, USA }

\author[0000-0002-5529-6780]{R.~Babu}
\affiliation{Department of Physics, Michigan Technological University, Houghton, MI, USA }

\author[0000-0003-3207-105X]{E.~Belmont-Moreno}
\affiliation{Instituto de F\'{i}sica, Universidad Nacional Autónoma de México, Ciudad de Mexico, Mexico }

\author[0000-0002-4042-3855]{K.S.~Caballero-Mora}
\affiliation{Universidad Autónoma de Chiapas, Tuxtla Gutiérrez, Chiapas, México}

\author[0000-0003-2158-2292]{T.~Capistrán}
\affiliation{Instituto de Astronom\'{i}a, Universidad Nacional Autónoma de México, Ciudad de Mexico, Mexico }

\author[0000-0002-8553-3302]{A.~Carramiñana}
\affiliation{Instituto Nacional de Astrof\'{i}sica, Óptica y Electrónica, Puebla, Mexico }

\author[0000-0002-6144-9122]{S.~Casanova}
\affiliation{Institute of Nuclear Physics Polish Academy of Sciences, PL-31342 IFJ-PAN, Krakow, Poland }

\author{O.~Chaparro-Amaro}
\affiliation{Centro de Investigaci\'on en Computaci\'on, Instituto Polit\'ecnico Nacional, M\'exico City, M\'exico.}

\author[0000-0002-7607-9582]{U.~Cotti}
\affiliation{Universidad Michoacana de San Nicolás de Hidalgo, Morelia, Mexico }

\author[0000-0002-1132-871X]{J.~Cotzomi}
\affiliation{Facultad de Ciencias F\'{i}sico Matemáticas, Benemérita Universidad Autónoma de Puebla, Puebla, Mexico }

\author[0000-0002-7747-754X]{S.~Coutiño de León}
\affiliation{Department of Physics, University of Wisconsin-Madison, Madison, WI, USA }

\author[0000-0001-9643-4134]{E.~De la Fuente}
\affiliation{Departamento de F\'{i}sica, Centro Universitario de Ciencias Exactas e Ingenierias, Universidad de Guadalajara, Guadalajara, Mexico }

\author{R.~Diaz Hernandez}
\affiliation{Instituto Nacional de Astrof\'{i}sica, Óptica y Electrónica, Puebla, Mexico }

\author[0000-0002-2987-9691]{M.A.~DuVernois}
\affiliation{Department of Physics, University of Wisconsin-Madison, Madison, WI, USA }

\author[0000-0003-2169-0306]{M.~Durocher}
\affiliation{Physics Division, Los Alamos National Laboratory, Los Alamos, NM, USA }

\author[0000-0002-0087-0693]{J.C.~Díaz-Vélez}
\affiliation{Departamento de F\'{i}sica, Centro Universitario de Ciencias Exactas e Ingenierias, Universidad de Guadalajara, Guadalajara, Mexico }

\author[0000-0001-5737-1820]{K.~Engel}
\affiliation{Department of Physics, University of Maryland, College Park, MD, USA }

\author[0000-0001-7074-1726]{C.~Espinoza}
\affiliation{Instituto de F\'{i}sica, Universidad Nacional Autónoma de México, Ciudad de Mexico, Mexico }

\author[0000-0002-8246-4751]{K.L.~Fan}
\affiliation{Department of Physics, University of Maryland, College Park, MD, USA }

\author[0000-0002-6305-3009]{M.~Fernández Alonso}
\affiliation{Department of Physics, Pennsylvania State University, University Park, PA, USA }
\affiliation{Gran Sasso Science Institute (GSSI), Via Iacobucci 2, I-67100 L’Aquila, Italy }

\author[0000-0002-0173-6453]{N.~Fraija}
\affiliation{Instituto de Astronom\'{i}a, Universidad Nacional Autónoma de México, Ciudad de Mexico, Mexico }

\author[0000-0002-4188-5584]{J.A.~García-González}
\affiliation{Tecnologico de Monterrey, Escuela de Ingenier\'{i}a y Ciencias, Ave. Eugenio Garza Sada 2501, Monterrey, N.L., Mexico, 64849}
\author[0000-0003-1122-4168]{F.~Garfias}
\affiliation{Instituto de Astronom\'{i}a, Universidad Nacional Autónoma de México, Ciudad de Mexico, Mexico }
\author[0000-0002-5209-5641]{M.M.~González}
\affiliation{Instituto de Astronom\'{i}a, Universidad Nacional Autónoma de México, Ciudad de Mexico, Mexico }
\author[0000-0002-9790-1299]{J.A.~Goodman}
\affiliation{Department of Physics, University of Maryland, College Park, MD, USA }

\author[0000-0001-9844-2648]{J.P.~Harding}
\affiliation{Physics Division, Los Alamos National Laboratory, Los Alamos, NM, USA }

\author[0000-0002-2565-8365]{S.~Hernandez}
\affiliation{Instituto de F\'{i}sica, Universidad Nacional Autónoma de México, Ciudad de Mexico, Mexico }
\author[0000-0002-5447-1786]{D.~Huang}
\affiliation{Department of Physics, Michigan Technological University, Houghton, MI, USA }
\author[0000-0002-5527-7141]{F.~Hueyotl-Zahuantitla}
\affiliation{Universidad Autónoma de Chiapas, Tuxtla Gutiérrez, Chiapas, México}
\author{P.~Hüntemeyer}
\affiliation{Department of Physics, Michigan Technological University, Houghton, MI, USA }
\author[0000-0001-5811-5167]{A.~Iriarte}
\affiliation{Instituto de Astronom\'{i}a, Universidad Nacional Autónoma de México, Ciudad de Mexico, Mexico }

\author[0000-0002-6738-9351]{A.~Jardin-Blicq}
\affiliation{Max-Planck Institute for Nuclear Physics, 69117 Heidelberg, Germany}
\affiliation{Department of Physics, Faculty of Science, Chulalongkorn University, 254 Phayathai Road, Pathumwan, Bangkok 10330, Thailand}
\affiliation{National Astronomical Research Institute of Thailand (Public Organization), Don Kaeo, MaeRim, Chiang Mai 50180, Thailand}

\author[0000-0003-4467-3621]{V.~Joshi}
\affiliation{Erlangen Centre for Astroparticle Physics, Friedrich-Alexander-Universit\"at Erlangen-N\"urnberg, Erlangen, Germany}
\author[0000-0003-4785-0101]{D.~Kieda}
\affiliation{Department of Physics and Astronomy, University of Utah, Salt Lake City, UT, USA }
\author[0000-0002-2467-5673]{W.H.~Lee}
\affiliation{Instituto de Astronom\'{i}a, Universidad Nacional Autónoma de México, Ciudad de Mexico, Mexico }
\author[0000-0003-2696-947X]{J.T.~Linnemann}
\affiliation{Department of Physics and Astronomy, Michigan State University, East Lansing, MI, USA }
\author[0000-0001-8825-3624]{A.L.~Longinotti}
\affiliation{Instituto de Astronom\'{i}a, Universidad Nacional Autónoma de México, Ciudad de Mexico, Mexico }
\author[0000-0003-2810-4867]{G.~Luis-Raya}
\affiliation{Universidad Politecnica de Pachuca, Pachuca, Hgo, Mexico }
\author[0000-0001-8088-400X]{K.~Malone}
\affiliation{Physics Division, Los Alamos National Laboratory, Los Alamos, NM, USA }
\author[0000-0001-9052-856X]{O.~Martinez}
\affiliation{Facultad de Ciencias F\'{i}sico Matemáticas, Benemérita Universidad Autónoma de Puebla, Puebla, Mexico }
\author[0000-0002-2824-3544]{J.~Martínez-Castro}
\affiliation{Centro de Investigaci\'on en Computaci\'on, Instituto Polit\'ecnico Nacional, M\'exico City, M\'exico.}
\author[0000-0002-2610-863X]{J.A.~Matthews}
\affiliation{Dept of Physics and Astronomy, University of New Mexico, Albuquerque, NM, USA }
\author[0000-0002-8390-9011]{P.~Miranda-Romagnoli}
\affiliation{Universidad Autónoma del Estado de Hidalgo, Pachuca, Mexico }
\author[0000-0002-1114-2640]{E.~Moreno}
\affiliation{Facultad de Ciencias F\'{i}sico Matemáticas, Benemérita Universidad Autónoma de Puebla, Puebla, Mexico }
\author[0000-0002-7675-4656]{M.~Mostafá}
\affiliation{Department of Physics, Pennsylvania State University, University Park, PA, USA }
\author[0000-0003-0587-4324]{A.~Nayerhoda}
\affiliation{Institute of Nuclear Physics Polish Academy of Sciences, PL-31342 IFJ-PAN, Krakow, Poland }
\author[0000-0003-1059-8731]{L.~Nellen}
\affiliation{Instituto de Ciencias Nucleares, Universidad Nacional Autónoma de Mexico, Ciudad de Mexico, Mexico }
\author[0000-0001-9428-7572]{M.~Newbold}
\affiliation{Department of Physics and Astronomy, University of Utah, Salt Lake City, UT, USA }
\author[0000-0002-6859-3944]{M.U.~Nisa}
\affiliation{Department of Physics and Astronomy, Michigan State University, East Lansing, MI, USA }
\author[0000-0001-7099-108X]{R.~Noriega-Papaqui}
\affiliation{Universidad Autónoma del Estado de Hidalgo, Pachuca, Mexico }
\author[0000-0002-5448-7577]{N.~Omodei}
\affiliation{Department of Physics, Stanford University: Stanford, CA 94305–4060, USA}
\author{A.~Peisker}
\affiliation{Department of Physics and Astronomy, Michigan State University, East Lansing, MI, USA }
\author[0000-0002-8774-8147]{Y.~Pérez Araujo}
\affiliation{Instituto de Astronom\'{i}a, Universidad Nacional Autónoma de México, Ciudad de Mexico, Mexico }
\author[0000-0001-5998-4938]{E.G.~Pérez-Pérez}
\affiliation{Universidad Politecnica de Pachuca, Pachuca, Hgo, Mexico }
\author[0000-0002-6524-9769]{C.D.~Rho}
\affiliation{University of Seoul, Seoul, Rep. of Korea}
\author[0000-0003-1327-0838]{D.~Rosa-González}
\affiliation{Instituto Nacional de Astrof\'{i}sica, Óptica y Electrónica, Puebla, Mexico }
\author{H.~Salazar}
\affiliation{Facultad de Ciencias F\'{i}sico Matemáticas, Benemérita Universidad Autónoma de Puebla, Puebla, Mexico }
\author[0000-0002-9312-9684]{D.~Salazar-Gallegos}
\affiliation{Department of Physics and Astronomy, Michigan State University, East Lansing, MI, USA }
\author[0000-0002-8610-8703]{F.~Salesa Greus}
\affiliation{Institute of Nuclear Physics Polish Academy of Sciences, PL-31342 IFJ-PAN, Krakow, Poland }
\affiliation{Instituto de Física Corpuscular, CSIC, Universitat de València, E-46980, Paterna, Valencia, Spain}
\author[0000-0001-6079-2722]{A.~Sandoval}
\affiliation{Instituto de F\'{i}sica, Universidad Nacional Autónoma de México, Ciudad de Mexico, Mexico }
\author{J.~Serna-Franco}
\affiliation{Instituto de F\'{i}sica, Universidad Nacional Autónoma de México, Ciudad de Mexico, Mexico }
\author[0000-0002-1012-0431]{A.J.~Smith}
\affiliation{Department of Physics, University of Maryland, College Park, MD, USA }
\author{Y.~Son}
\affiliation{University of Seoul, Seoul, Rep. of Korea}
\author[0000-0002-1492-0380]{R.W.~Springer}
\affiliation{Department of Physics and Astronomy, University of Utah, Salt Lake City, UT, USA }
\author{O.~Tibolla}
\affiliation{Universidad Politecnica de Pachuca, Pachuca, Hgo, Mexico }
\author[0000-0001-9725-1479]{K.~Tollefson}
\affiliation{Department of Physics and Astronomy, Michigan State University, East Lansing, MI, USA }
\author[0000-0002-1689-3945]{I.~Torres}
\affiliation{Instituto Nacional de Astrof\'{i}sica, Óptica y Electrónica, Puebla, Mexico }
\author{R.~Torres-Escobedo}
\affiliation{Tsung Dao Lee Institute and School of Physics and Astronomy, Shanghai Jiao Tong University, Shanghai, China}
\author[0000-0003-1068-6707]{R.~Turner}
\affiliation{Department of Physics, Michigan Technological University, Houghton, MI, USA }
\author[0000-0002-2748-2527]{F.~Ureña-Mena}
\affiliation{Instituto Nacional de Astrof\'{i}sica, Óptica y Electrónica, Puebla, Mexico }
\author[0000-0001-6876-2800]{L.~Villaseñor}
\affiliation{Facultad de Ciencias F\'{i}sico Matemáticas, Benemérita Universidad Autónoma de Puebla, Puebla, Mexico }
\author[0000-0001-6798-353X]{X.~Wang}
\affiliation{Department of Physics, Michigan Technological University, Houghton, MI, USA }

\author{T.~Weisgarber}
\affiliation{Department of Physics, University of Wisconsin-Madison, Madison, WI, USA }
\author[0000-0002-6623-0277]{E.~Willox}
\affiliation{Department of Physics, University of Maryland, College Park, MD, USA }
\author[0000-0003-0513-3841]{H.~Zhou}
\author{C.~de León}
\affiliation{Universidad Michoacana de San Nicolás de Hidalgo, Morelia, Mexico }

\collaboration{80}{HAWC Collaboration}

\begin{abstract}
    The extragalactic background light (EBL) contains all the radiation emitted by nuclear and accretion processes in stars and compact objects since the epoch of recombination. Measuring the EBL density directly is challenging, especially in the near- to far-infrared waveband, mainly due to the zodiacal light foreground. Instead, gamma-ray astronomy offers the possibility to indirectly set limits on the EBL by studying the effects of gamma-ray absorption in the very high energy (VHE:$>$100 GeV) spectra of distant blazars. The High Altitude Water Cherenkov gamma ray observatory (HAWC) is one of the few instruments sensitive to gamma rays with energies above 10 TeV. This offers the opportunity to probe the EBL in the near/mid IR region: $\lambda$ = 1 $\mu$m - 100 $\mu$m. In this study, we fit physically motivated emission models to \textit{Fermi Large Area Telescope (LAT)} GeV data to extrapolate the intrinsic TeV spectra of blazars. We then simulate a large number of absorbed spectra for different randomly generated EBL model shapes and calculate Bayesian credible bands in the EBL intensity space by comparing and testing the agreement between the absorbed spectra and HAWC extragalactic observations of two blazars. The resulting bands are in agreement with current EBL lower and upper limits, showing a downward trend towards higher wavelength values $\lambda>10 \mu$m also observed in previous measurements.
\end{abstract}

\section{Introduction}
\label{sec:Intro}
The extragalactic background light (EBL) comprises all radiation released by nuclear and accretion processes since the epoch of recombination. It consists of all emitted radiation from stars and compact object surroundings, including that absorbed/re-emitted by dust and accumulated over all redshifts. Measuring and constraining  this background radiation is crucial  to understand star formation processes and galaxy evolution models. Our current knowledge of the EBL is limited, its direct measurements are challenging due to foreground contamination coming, mainly, from the zodiacal light. However, upper and lower limits have been established using various methods, e.g. integrated galaxy counts from optical observations with the Hubble Space Telescope (\citealt{2000ApJ...542L..79G};  \citealt{2000MNRAS.312L...9M}) and  infrared (IR) observations using the Spitzer Space Telescope  (\citealt{2004ApJS..154...39F}; \citealt{2004ApJS..154...70P}) and the Infrared Space Observatory \citep{2002A&A...384..848E}. An extensive discussion on EBL related matters can be found in \cite{2002A&A...384..848E}, \cite{2011ApJ...733...77O} and \cite{2016_EBL_review}. 

Over the past two decades, VHE gamma-ray observations have been used to help constrain the spectral properties of the EBL, particularly with observations from blazars, a sub-type of Active Galactic Nuclei (AGN) with ultra-relativistic jets oriented close to the observer's line of sight.  VHE gamma rays coming from  blazars interact via pair production with the EBL photons (\citealt{Nikishov1962JPSJS..17C.175G}; \citealt{PhysRev.155.1404}) producing imprints in the observed energy spectra of distant sources. These imprints, along with intrinsic spectral assumptions, can be used to derive limits on the EBL near/mid IR range using VHE observations of blazars (e.g. \citealt{Aharonian_2007}; \citealt{2007A&A...471..439M}; \citealt{2011ApJ...733...77O}; \citealt{Biteau_2015}; \citealt{2017A&A...606A..59H}; \citealt{Abeysekara_2019_VERITAS}; \citealt{2019MNRAS.486.4233A}).
The gamma-ray horizon establishes the energy at which the intensity of radiation is diminished by a factor 1/$e$ with respect to the emitted intensity. This energy value depends on the distance of the source and it needs to be taken into account when selecting suitable candidate sources for this type of study. If the source is too close, the absorption effect will only be measurable at higher energies, where, depending on the detector and the energy of the gamma ray, the sensitivity might be too low. On the other hand, distant sources will be dimmer, precisely due to the EBL absorption, so these are not ideal candidates to  work with either \citep{Franceschini_2019_IR_IdeallSource}. Equation \ref{eq:EBL_Range} approximately relates the energy of a gamma ray ($E_\gamma$) with the wavelength range of the EBL radiation ($\lambda_{EBL}$) involved in the pair production interaction:

\begin{equation}
\lambda_{EBL} \sim \left[0.5\mu m - 5\mu m\right] \times \left(\frac{E_{\gamma}}{1 \text{ TeV}}\right) \times \left(1+z\right)^2.
\label{eq:EBL_Range}
\end{equation}
where $z$ is the redshift of the source emitting the gamma ray. This equation is useful to estimate the EBL probing power when considering a specific source observed with a specific instrument. 

The High Altitude Water Cherenkov gamma ray observatory (HAWC) is a water Cherenkov detector that has been operational since 2015 (further technical details can be found in section \ref{sec:HAWC}), and that has detected extragalactic sources significantly up to 10 TeV \citep{albert2021longterm}. This energy is close to the one established by the gamma-ray horizon for sources like the two blazars Markarian 501 and Markarian 421 (from now on referred to as \textit{Markarians}), putting HAWC in an advantageous position to potentially probe the mid-IR region of the EBL using observations from these sources.


In this study, physically motivated emission models and data from the \textit{Fermi Large Area Telescope (Fermi-LAT)} are used to construct an intrinsic spectrum for each of the Markarians (see section \ref{sec:IntSpec}). A large number of randomly generated EBL model shapes are used to apply the EBL attenuation effect to these intrinsic spectra to compare with HAWC data (see section \ref{sec:ebl_models}). The comparison is performed using \textit{threeML}\footnote{\url{https://threeml.readthedocs.io/en/stable/index.html}}, a software package for likelihood fitting, in a way that each EBL model can be assigned a likelihood value that expresses the agreement between that particular spectral realization and HAWC data (see section \ref{sec:3ML}). This method has the advantage of being independent of any particular EBL shape and of assuming physically motivated intrinsic spectral properties for the sources. Finally, weights are applied to each model in accordance to their calculated likelihood value, and credible intervals in the EBL spectral energy distribution (SED) space are derived (section \ref{sec:Results}).

\section{Data: HAWC \& \textit{Fermi-LAT}}
\label{sec:HAWC}

HAWC is an array of 300 water Cherenkov detector tanks located in Sierra Negra, Mexico, at an altitude of 4100 m above the sea level and covering a total area of 22000 m$^2$.
Each tank has four photo-multiplier tubes (PMTs) facing upwards that can detect the Cherenkov light in the water from the transit of secondary particles, which are produced by gamma rays and cosmic rays interacting with the atmosphere.  HAWC triggers with a rate of 25 kHz and  has  a  duty  cycle  of $>$95\%. The observatory continuously monitors 2/3 of the sky, detecting gamma rays in the energy range between 100 GeV and several hundred TeV.
More information on the HAWC observatory operation, performance and the way air shower event data are reconstructed can be found in \cite{Abeysekara_2017}.
\label{sec:mrks}
For this analysis, specific \textit{Fermi-LAT} and HAWC data from the blazars Markarian 421 ($z=0.031$) and Markarian 501 ($z=0.034$) were selected (redshift sourced from NED\footnote{The NASA/IPAC Extragalactic Database (NED)
is operated by the Jet Propulsion Laboratory, California Institute of Technology,
under contract with the National Aeronautics and Space Administration.}). These are two extensively studied  extragalactic sources and the brightest in the TeV band. Both sources have been significantly detected by HAWC above 300 GeV up to 10 TeV \citep{albert2021longterm} and by \textit{Fermi-LAT} between 100 MeV and 1 TeV (\citealt{Abdo_2011_421}; \citealt{Abdo_2011_501}).
For the HAWC dataset, 1343 days of data were used, taken between June 2015 and June 2019 \citep{Abeysekara_2019}. The analysis was performed using maps created with a special algorithm for reconstructing and determining the energy of the primary gamma rays. This algorithm, denominated ground parameter algorithm, is based primarily on the charge density at a fixed optimal distance from the shower axis and it divides the energy range into quarter-decade bins in log10($E$), beginning at log($E$/TeV)=-0.5 (0.316 TeV) and ending at log($E$/TeV)=2.5 (316 TeV).
The ground parameter energy estimator was chosen because it is optimal for higher energies, between 10 TeV and 316 TeV \citep{Abeysekara_2019}, where the instrument could potentially probe the EBL mid-IR region. 
For the \textit{Fermi-LAT} dataset, a time period corresponding to the HAWC dataset was selected between 57180 - 58640 MJD. This dataset is within the energy range of 100 MeV - 316 GeV, where the absorption is, at most, around 5\% for the Markarians redshift\footnote{estimated assuming the \citet{2011MNRAS.410.2556D} EBL model}. The \textit{Fermi-LAT} analysis is described in more detail in section \ref{sec:IntSpec}.

\section{EBL Analysis}
\label{sec:method}
The approach adopted in this study consists of calculating EBL intensity limits by comparing the expected absorbed flux with actual HAWC observations. The method is similar to previous EBL studies: \cite{2007A&A...471..439M}, \cite{2011ApJ...733...77O}, \cite{Biteau_2015} and, in particular, to \cite{Abeysekara_2019_VERITAS}, presented by the VERITAS collaboration. In the latter analysis, a large number of random generated EBL models is used to calculate the corresponding de-absorbed spectra of selected blazars, and then each model is weighted using criteria based on intrinsic assumptions for these sources to finally derive limits to the EBL intensity. In the present analysis, the EBL models are used to simulate the absorption effect which is then applied over the assumed intrinsic spectra. The resulting absorbed spectra are then compared to the real HAWC data by calculating a likelihood value. Finally, each EBL model is weighted according to their corresponding likelihood value.
 
\subsection{Intrinsic Spectra: \textit{Fermi-LAT} \& Naima}
\label{sec:IntSpec}
 A possible and reasonable assumption is to consider the intrinsic TeV spectrum of a given source as an extension of a physical emission model that is in agreement with GeV observations. This assumption relies on the fact that gamma rays in the high energy regime (HE: 100 MeV - 100 GeV) and relatively low redshifts ($z\lesssim0.1$) are not significantly affected by EBL absorption, and therefore the observed spectrum can be safely considered to be practically the same as the intrinsic one (e.g. \citealt{2010ApJ...716...30A, 2011ApJ...733...77O, 2013ApJ...768L..31F}). In the present study, a \textit{Fermi-LAT} standard fitting analysis was performed to obtain spectral flux points corresponding to Markarian 421 and Markarian 501, using observations from a similar time period as the one observed by HAWC (for details see section \ref{sec:mrks}). The \textit{Fermi-LAT} analysis was carried out using the instrument response function P8R2\_SOURCE\_V6, and the spectral parameters are estimated by the binned maximum likelihood method using the \textit{Fermipy v.0.17.3} package. The analysis was also performed with a zenith cut of 90$^o$ and, as mentioned in section \ref{sec:mrks}, within the energy range of 100 MeV - 316 GeV to minimize possible absorption effects while including as much data as possible. The events were extracted within a 10 degree region of interest (RoI) centered on each source position. The background model includes sources from the Preliminary \textit{Fermi-LAT} 8-year point sources catalog\footnote{\url{https://fermi.gsfc.nasa.gov/ssc/data/access/lat/fl8y/}}, Galactic diffuse emission \textit{glliemv06.fits} and the isotropic diffuse emission \textit{isoP8R2SOURCEV6v06.txt} models. 
 
 To avoid the overestimation of the intrinsic flux, each source lowest energy fluxpoint observed by HAWC \citep{albert2021longterm} is used as a guide; observed Mrk\,421 and Mrk\,501 fluxpoints at an energy of 830 GeV  and 750 GeV, respectively, are de-asbsorbed according to \citet{Franceschini} model. These de-absorbed fluxpoints are included in the set of flux points obtained from the \textit{Fermi-LAT} analysis described above, and altogether are used to fit a physically motivated emission model with \textit{Naima}, a Python software software that calculates the non-thermal emission from a leptonic or hadronic population of particles \citep{zabalza2015naima}.
 
 In this case, the dataset is fit with a Synchrotron-Self-Compton (SSC) model, a scenario in which synchrotron radiation is produced by electrons moving at relativistic velocities in randomly oriented magnetic fields and up-scattered by inverse Compton into higher energies by the same electron population. The leptonic distribution is chosen to follow an exponentially cut-off powerlaw (ECPL) distribution:
 
 \begin{equation}
    f\left(E\right) = A\left(\frac{E}{E}_{0}\right)^{-\alpha} \exp\left(-\frac{E}{E_{c}}\right).
\end{equation}
 where $A$ is a normalization factor, $E_0$ is a reference energy set at 1 TeV, $\alpha$ is the power law index and $E_c$ the cut-off energy.
 The decision for this particular leptonic distribution is motivated by other studies performed on the Markarians by HAWC \citep{albert2021longterm} and others  (\citealt{Mrk421_SSC}; \citealt{Abdo_2011_501}). 
 Based on the results reported in \cite{albert2021longterm}, the magnetic field and emission radius values are set at 0.03 G and 4 $\times 10^{16}$cm  for Mrk\,421 and 0.02 G and 1 $\times 10^{17}$cm for Mrk\,501, respectively. Table \ref{tab:Naima_param} shows the best fit parameters for the electron distributions of each source. The best fit models are then extrapolated into TeV energies (see figure \ref{fig:Int_Mrk421}) and used as templates for the intrinsic spectra of the sources (see section \ref{sec:3ML}).
 
\begin{table*}
\caption{Electron distribution parameters resulting from the SSC emission model likelihood fit performed with NAIMA to \textit{Fermi-LAT} data from Mrk\,421 and Mrk\,501. $A$ is a normalization factor, $E_0$ is a reference energy set at 1 TeV, $\alpha$ is the power law index and $E_c$ the cut-off energy}
\centering
\begin{tabular}{lcc}
\hline 
Parameter  & Mrk\,421 & Mrk\,501  \\

\hline

$\alpha$  &  2.77$\pm$0.02  &   2.75$\pm$0.04  \\
E$_c$ [$TeV$]  &  (1.2$\pm$0.9) $\times$ 10$^{1}$  &  (1.85$\pm$0.75) $\times$ 10$^1$    \\
A  [1/eV] &  (1.4$\pm$0.1) $\times$ 10$^{36}$  &  (0.9$\pm$0.1) $\times$ 10$^{36}$   \\

\hline 

\end{tabular}
\label{tab:Naima_param}
\end{table*}


\subsection{EBL models}
\label{sec:ebl_models}
\begin{figure}
     \centering
     \includegraphics[width=0.5\textwidth]{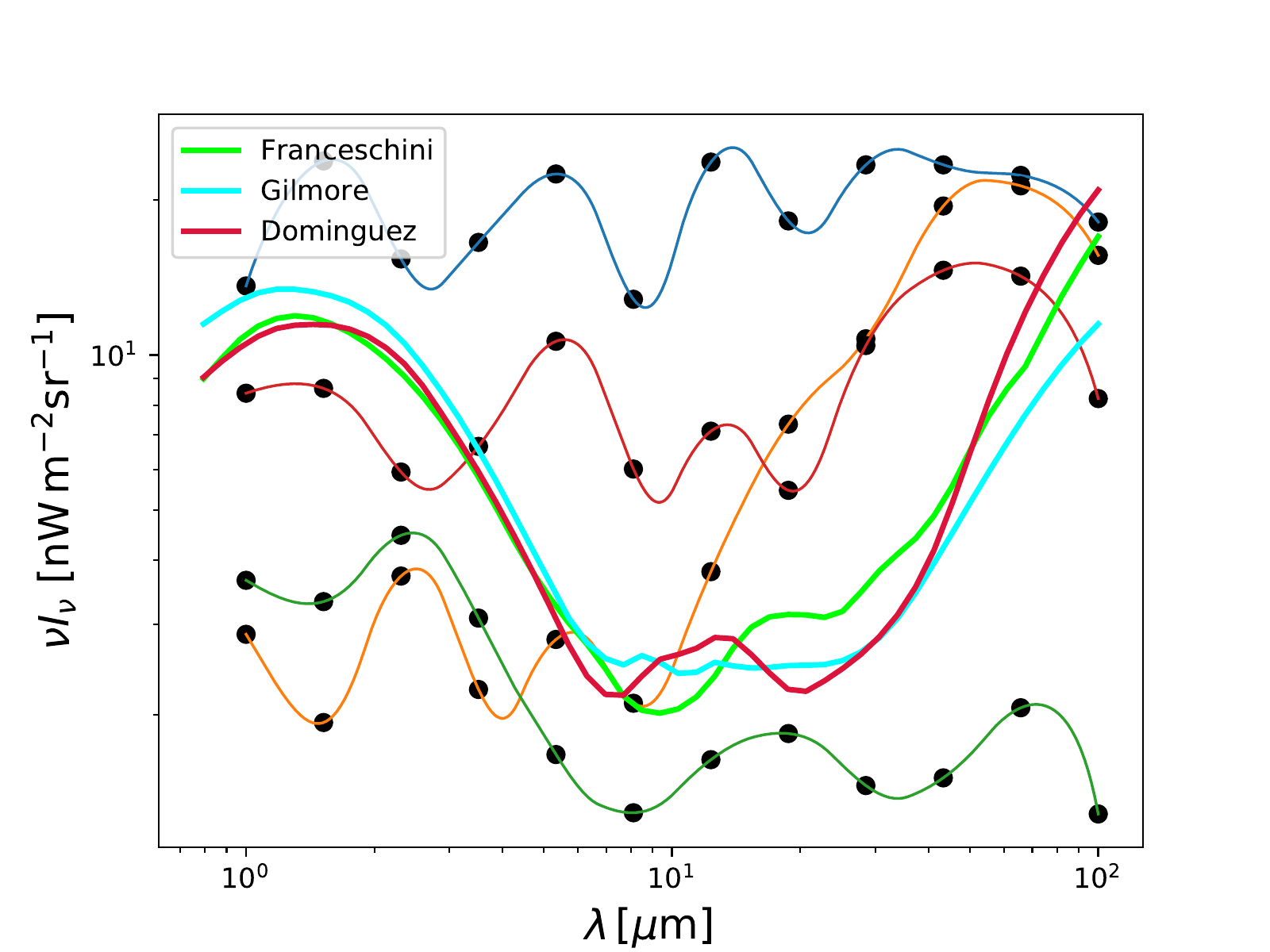}
     \caption{Sample of spline-interpolated models in the EBL spectral density space. Also shown, \cite{Franceschini}, \cite{Dominguez} and \cite{Gilmore} EBL models shapes.}
     \label{fig:splines}
 \end{figure}

To generate different EBL model shapes, a grid of 12 $\lambda$-values across the wavelength of the EBL photons between 1 $\mu$m and 100 $\mu$m is set. For each $\lambda$-value,an intensity (nW m$^{-2}$ sr$^{-1}$) value is assigned by randomly generating a number between 1 nW m$^{-2}$ sr$^{-1}$ and 50 nW m$^{-2}$ sr$^{-1}$, resulting in a flat initial distribution in EBL intensity; the range is chosen to contain the EBL upper and lower limits discussed in section \ref{sec:Intro}. A particular EBL shape is defined by interpolating each intensity point corresponding to each wavelength, using second order splines. To avoid sharp model shapes, intensity at consecutive grid points is not allowed to change by more than a factor of 2. For this reason and, since the interpolation is performed in increasing order in $\lambda$, intensities for higher $\lambda$ values are more conditionally sampled than lower $\lambda$ values. This bias, is then corrected by weighting the intensities in each grid point to recover the initial flat intensity distribution. Avoiding sharp models that are nonphysical has the caveat of neglecting possible features coming from polycyclic aromatic hydrocarbons, known to be present in the mid-IR range \citep{Lagache_2005}. However, for relatively small redshifts, like those considered in this analysis, these sharp features would be smoothed out to a bump \citep{Dominguez}.
Figure \ref{fig:splines} shows examples of EBL shapes generated in this way. A total of 30000 models are generated at redshift $z=0$. The EBL intensity can be then used to compute the optical depth $\tau(E,z)$, which quantifies the absorption effect of gamma rays with energy ($E$) and traveling a given distance associated with a redshift value $z$. The theoretical framework behind these calculations can be found in e.g. \cite{2013APh....43..112D}; \cite{Biteau_2015}; \cite{2017A&A...606A..59H}. In this analysis, the $\tau$ values are computed using the  \textit{ebltable} Python package\footnote{\url{https://github.com/me-manu/ebltable}} to read in and interpolate tables for EBL density and to calculate the resulting opacity for high energy gamma rays.
The $\tau$ values are computed between 0.1 and 50 TeV and for redshifts between 0.03-0.04, chosen respectively to include the Markarian's highest energy points measured by HAWC and their redshift. The points at which the calculations are done are evenly spaced in logarithmic space in energy, and linearly in redshift. A flat $\Lambda$CDM cosmology is assumed in the calculation, with dark-energy density $\Omega_\Lambda$ = 0.73, matter density $\Omega_M$= 0.27, and Hubble constant H$_0$=70 km s$^{-1}$ Mpc$^{-1}$.
The EBL number density $\eta_{\text{EBL}}$ is scaled with redshift as (1 + z)$^{3-f_{evo}}$ \citep{1996ApJ...456..124M}, and a value of $f_{evo}$=1.7 (evolution factor) is chosen following \cite{Abeysekara_2019_VERITAS}, where they find it to be consistent with the \cite{Franceschini}, \cite{Dominguez} and \cite{Gilmore} EBL models. 
The calculated opacity for each redshift-energy is then used to simulate the absorption process by applying the attenuation factor to the assumed intrinsic differential flux in the following way:
\begin{equation}
    \frac{dN}{dE}_{obs} = \frac{dN}{dE}_{int} \times e^{-\tau\left(E,z\right)}.
\end{equation}
where $\frac{dN}{dE}_{obs}$ is the observed differential flux and $\frac{dN}{dE}_{int}$ the assumed intrinsic differential flux. 

\begin{figure}[h]
     \centering
     \includegraphics[width=0.5\textwidth]{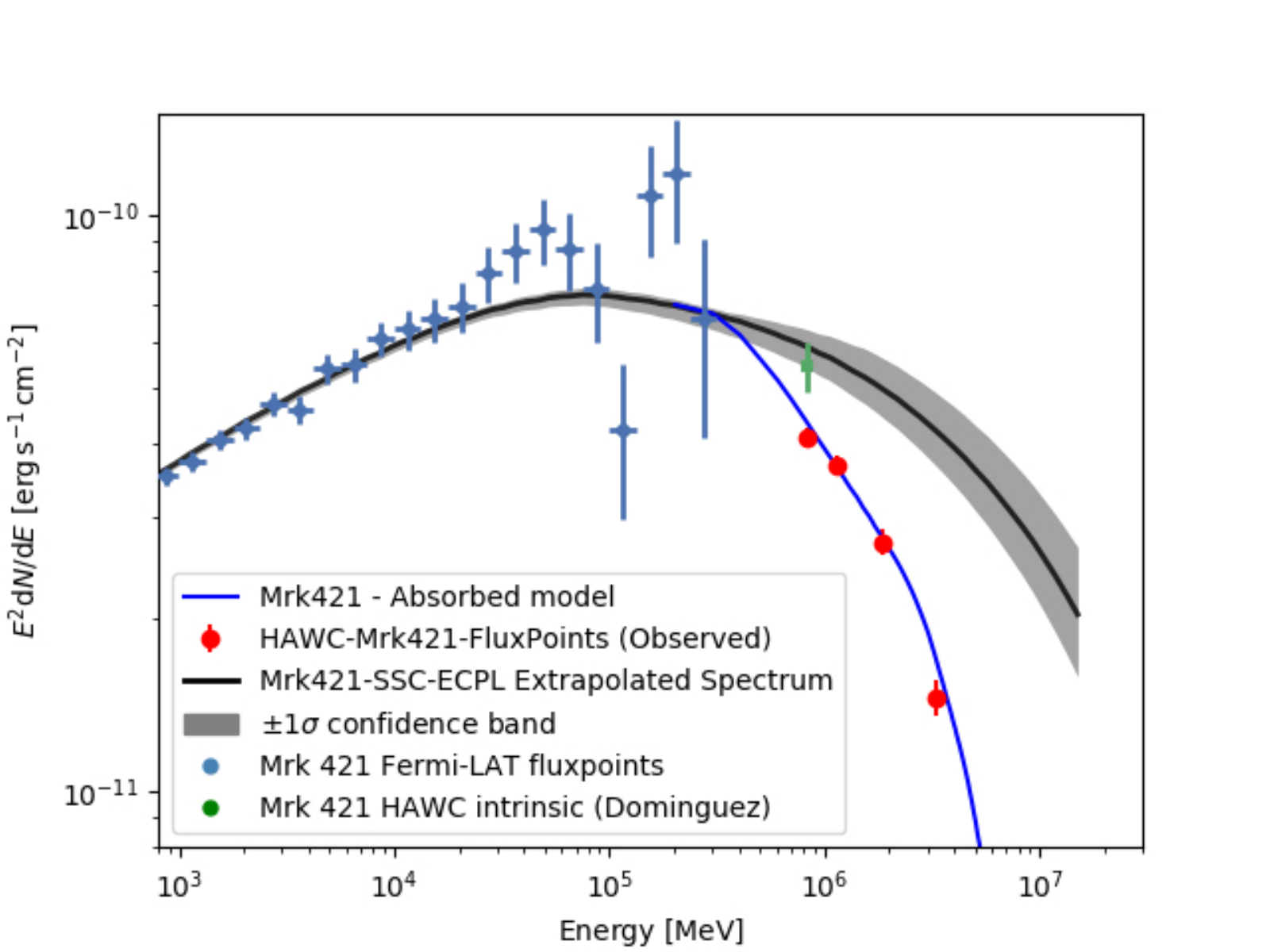}
     \caption{Extrapolated intrinsic emission spectrum for Mrk\,421 (black line) along with the $\pm$1$\sigma$ confidence band (statistical uncertainty only). Also shown, the resulting absorbed spectrum (blue line) according to a random EBL model along with HAWC data, the fluxpoints resulting from the $Fermi-LAT$ analysis (light-blue points) and the de-absorbed HAWC fluxpoint (green) according to \citet{Franceschini} EBL model.}
     \label{fig:Int_Mrk421}
 \end{figure}

Figure \ref{fig:Int_Mrk421} shows Mrk\,421 \textit{Fermi-LAT} flux points with its corresponding \textit{Naima-SSC} fit extrapolated into TeV energies. An example of the absorbed spectrum according to an EBL model along with HAWC data is also shown. 

\subsection{threeML framework, likelihood and limit extraction}
\label{sec:3ML}
The comparison between the expected absorbed model and HAWC data, is performed adopting a Bayesian approach and using the Multi-Mission Maximum Likelihood framework (\citealt{2015ICRC...34.1042V}; \citealt{2015ICRC...34..948Y}). This analysis pipeline is capable of handling data from a wide variety of astrophysical detectors. In this particular study the HAWC plugin (HAL\footnote{\url{https://github.com/threeML/hawc_hal}}) is used. The HAL fitting technique is based on a forward-folding method that assumes a spectral model shape for the source. In this case, a SSC source emission model is built using the \textit{astromodels} package, a useful framework to define models for likelihood or Bayesian analysis of astrophysical data\footnote{\url{https://astromodels.readthedocs.io/en/latest/}}. This emission model is customized for each source by plugging in the fit parameters obtained from the \textit{Naima} fit described in section \ref{sec:IntSpec}, to create an intrinsic spectrum that serves as an input for the threeML fitting pipeline. The EBL absorption is applied to the emission model while their spectral parameters are kept fixed. Finally, the forward-folding method, including detector response effects, is performed to fit the absorbed spectrum of the source using a maximum likelihood technique \citep{younk2015highlevel} to calculate a likelihood value for the fit ($\mathcal{L}$). The process is repeated for each of the 30000 EBL models generated as explained in section \ref{sec:ebl_models}. In this way, instead of optimizing the parameters of the source to find the ones that give the maximum likelihood, these are kept fixed, and only the EBL models are evaluated by their ability to reproduce the data. After the test, each EBL model is assigned a likelihood value that quantifies the agreement between that particular \textit{emission + absorption} model and the actual data. 
The maximum likelihood value corresponds to the EBL model whose simulated absorption best reproduces HAWC observations. Starting off from a prior flat intensity distribution for each wavelength, resulting from the uniform sampling described in section \ref{sec:ebl_models}, each model intensity is then assigned a weight as follows:
\begin{equation}
    W_i = \frac{\mathcal{L}_i}{\mathcal{L}_{max}}.
\end{equation}
where $\mathcal{L}_i$ is the likelihood value corresponding to the ``i-th" EBL model. The intensities corresponding to the \textit{maximum-likelihood} model are then assigned a weight of 1 and the rest of the model intensities are weighted by a factor of $\mathcal{L}_{i}$/$\mathcal{L}_{max}$, disfavoring EBL shapes that differ from the EBL model that best agrees with data. This approach is based on the idea of \textit{relative likelihood}, which is an estimate of the probability of a model to reproduce data, normalized by the maximum likelihood value \citep{StatisticalInference}.
When combining results from both sources, each EBL model is weighted by multiplying the individual source weights previously calculated: W$_{mrk421} \times $W$_{mrk501}$. For each $\lambda$-value, the \textit{maximum-likelihood} EBL models from each source are used to calculate an average intensity value for the combined result.

\begin{figure}
     \centering
     \includegraphics[width=0.5\textwidth]{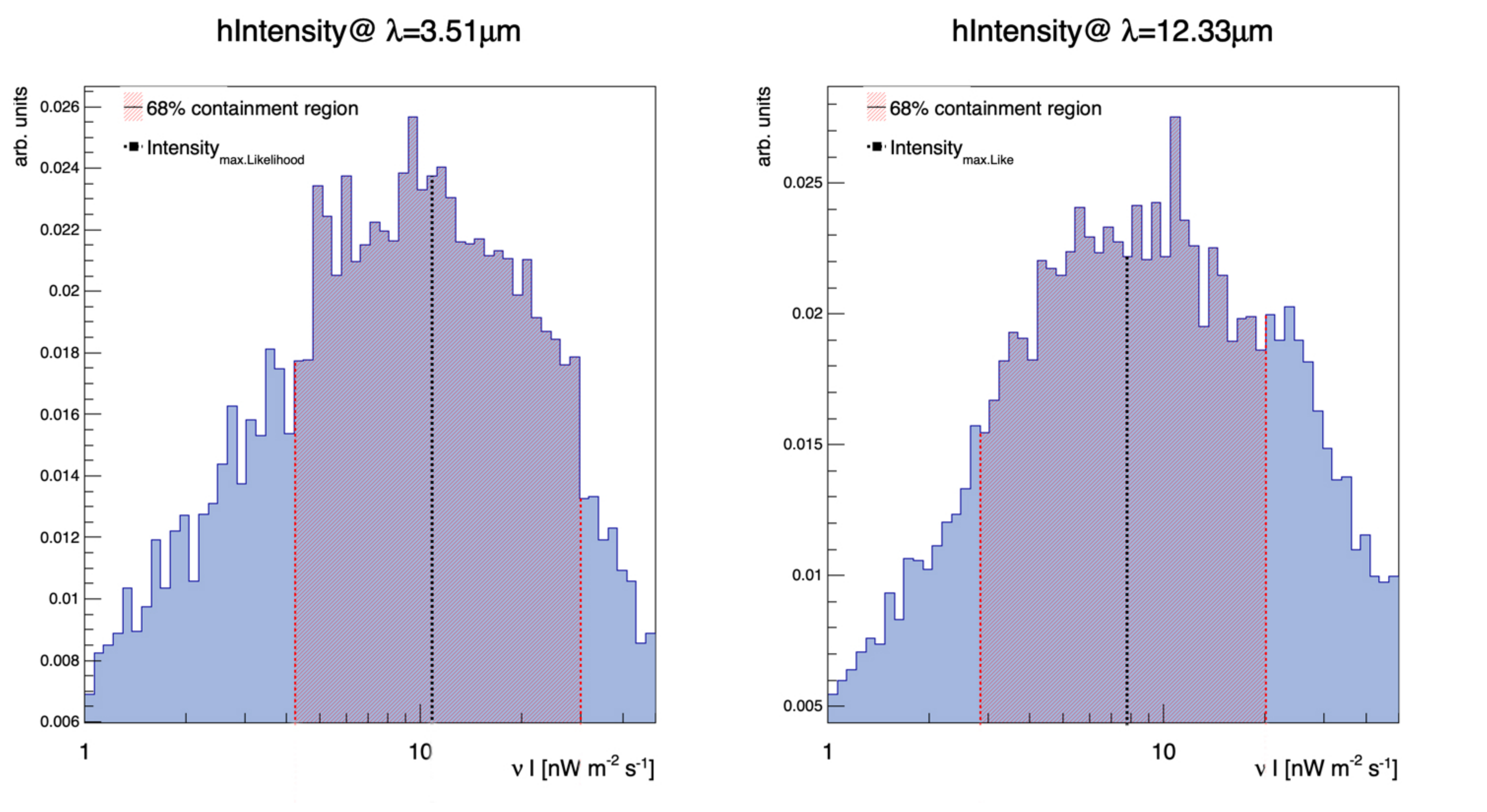}
     \caption{Weighted intensity distributions for $\lambda=3.5\mu$m and $\lambda=12.3\mu$m. The red hatching corresponds to the 68\% containment region. Black dashed line represents the intensity corresponding to the maximum likelihood model at that particular value of $\lambda$.}
     \label{fig:weight_int}
 \end{figure}

A projection of the EBL intensity probability for each value of $\lambda$ can be obtained from the histogram of weighted intensities, which represents the posterior distribution of intensities at that $\lambda$ given the observed data. Figure \ref{fig:weight_int} shows an example of these distributions for two values of $\lambda$. The distributions are initially (pre-weighting) flat on average, but fluctuate in accordance with the random sampling described in section \ref{sec:ebl_models}. This explains why the intensity corresponding to the \textit{maximum-likelihood} model (black-dashed line in figure \ref{fig:weight_int}), in spite of having the greatest weight, does not necessarily correspond with the maximum of the distribution.
 
 The credible intervals (analogous to confidence intervals in frequentist statistics) are obtained by integrating the 0.68 and 0.95 containment regions from the $\lambda$ value corresponding to the \textit{maximum-likelihood} model outwards. In this way, the 68\% and 95\% containment regions are defined around the model which best reproduces the data. For some wavelength values, the corresponding distributions are such that the limits fall outside considered range of intensity. For these cases, the containment region is reported without a lower/upper boundary. 
 To test the sensitivity of the method to the prior assumptions, the procedure is repeated varying the smoothing condition between consecutive knots and alternatively sampling a flat distribution in logarithm of the intensities instead of the original linear sampling (see section \ref{sec:ebl_models}). No significant changes are observed throughout the process with respect to the original prior conditions.

\section{Results and discussion}
\label{sec:Results}
\begin{figure*}[ht]
  \centering    
  \includegraphics[scale=0.5]{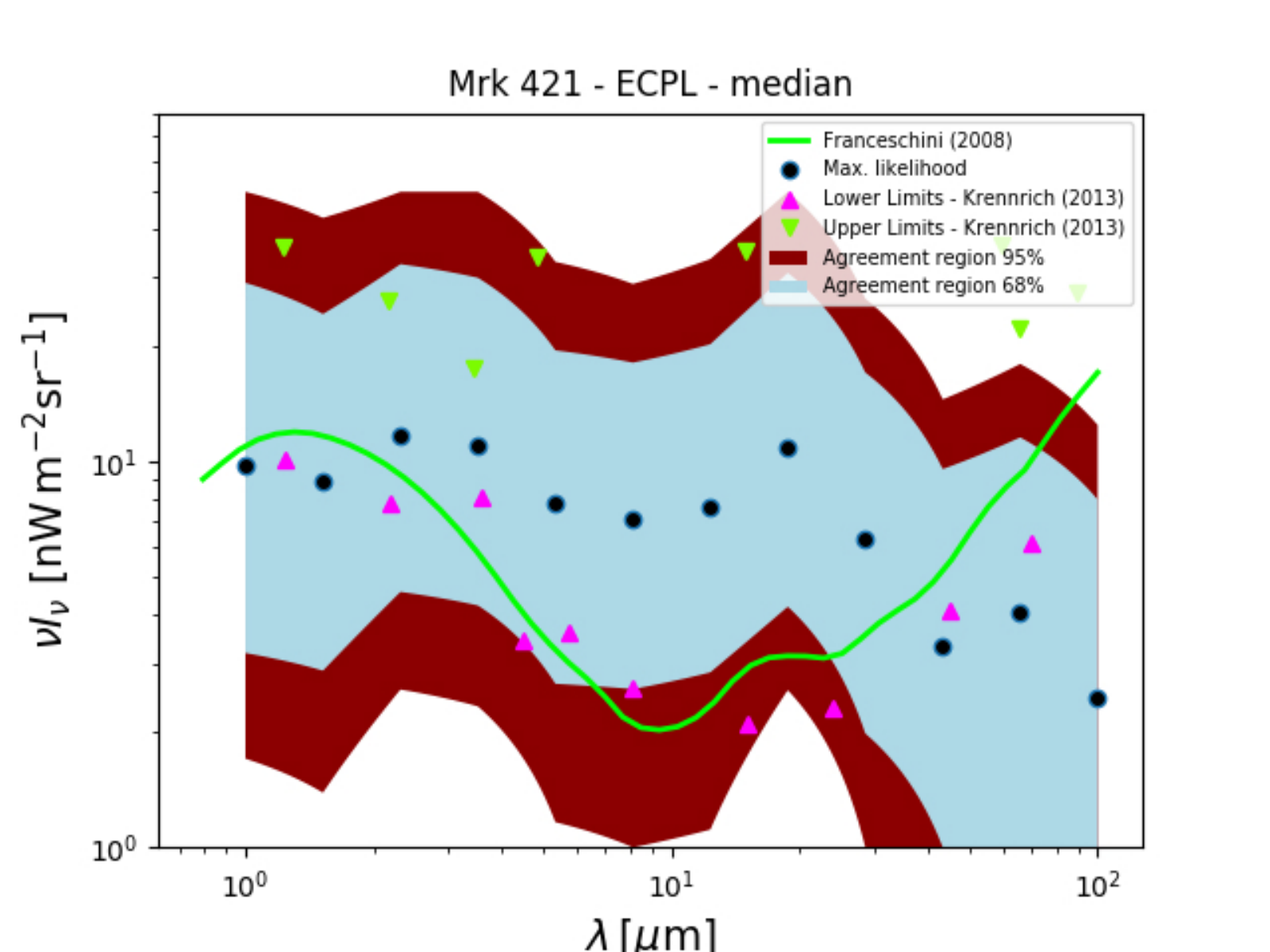}
  \includegraphics[scale=0.5]{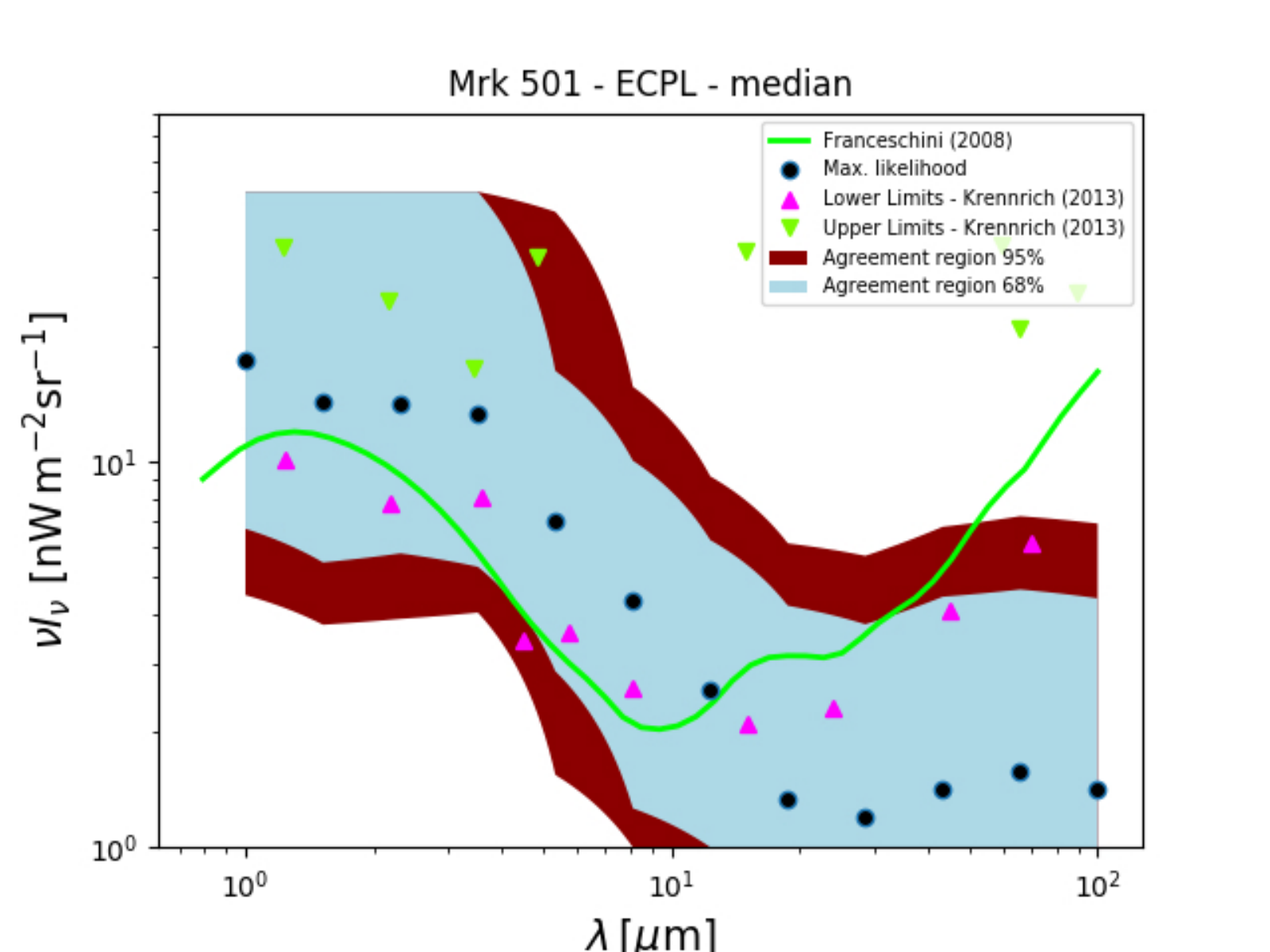}
  \caption{95\% and 68\% containment bands for the EBL intensity and different $\lambda$ values. Red circles correspond to the intensities of the model with the highest likelihood value. Also shown, lower limits from galaxy counts (cyan triangles), upper limits from direct measurements (green triangles) and \cite{Franceschini} EBL model for reference.}
  \label{fig:bands_sep}
\end{figure*}

The containment regions are calculated assuming SSC emission models from ECPL-leptonic distributions for both Mrk\,421 and Mrk\,501 intrinsic spectral models, as explained in section \ref{sec:IntSpec}. Figure \ref{fig:bands_sep} shows the resulting 68\% and 95\% containment regions for each of the considered sources as a function of wavelength.
The intensities corresponding to the model with the highest likelihood are shown in black circles, along with lower limits from galaxy counts and upper limits from direct measurements shown as upward and downward triangles, respectively.

\begin{figure*}[ht]
  \centering
  \includegraphics[scale=0.7]{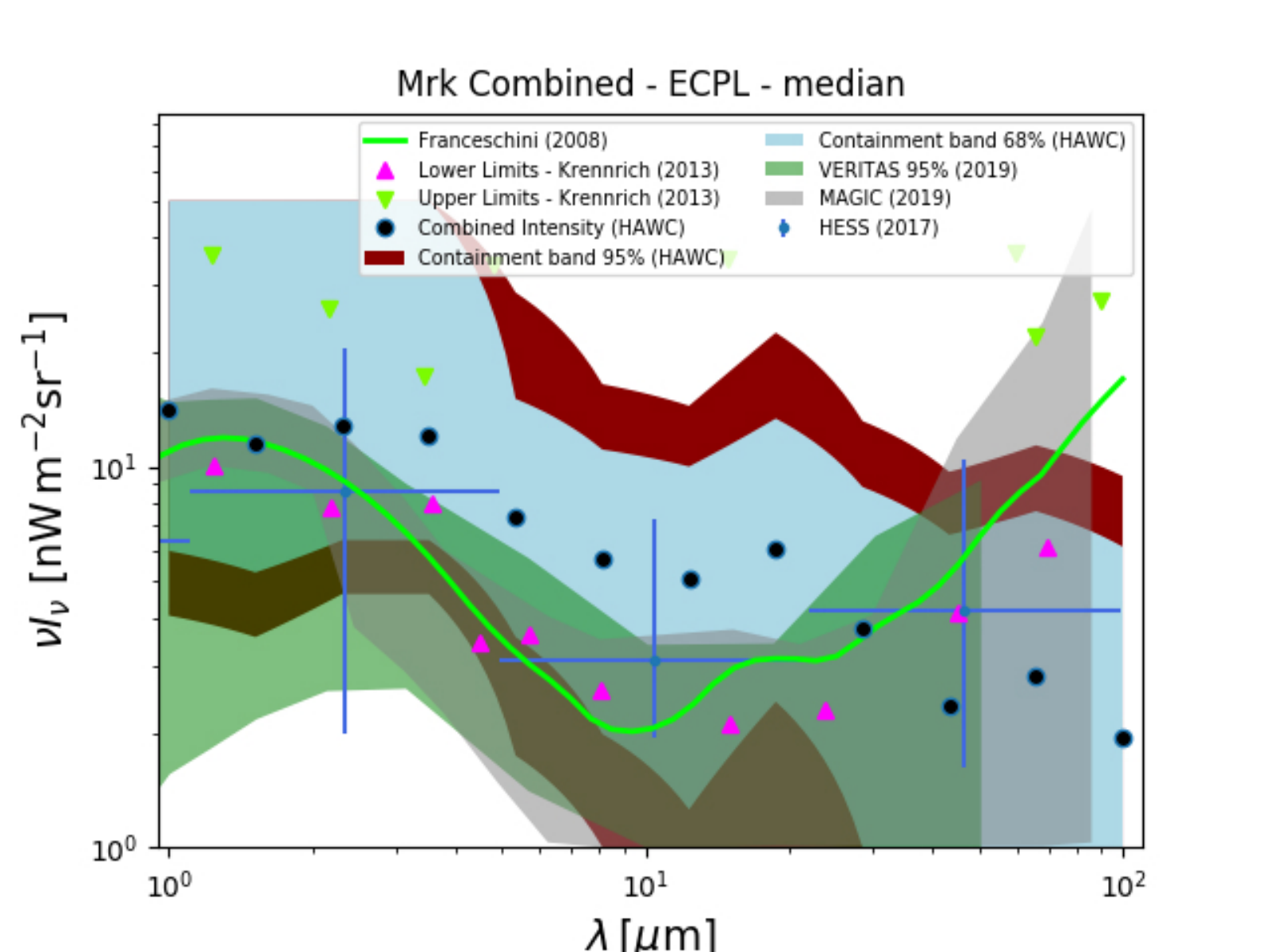}
  \caption{95\% and 68\% containment bands for the EBL intensity different $\lambda$ values for the combined results from Mrk\,421 and Mrk\,501. Red circles correspond to the combined intensities from each source highest likelihood models (see section \ref{sec:3ML}). Also shown, lower limits from galaxy counts (cyan triangles), upper limits from direct measurements (green triangles) and EBL measurements from VERITAS \citep{Abeysekara_2019_VERITAS}, MAGIC \citep{2019MNRAS.486.4233A} and HESS \citep{2017A&A...606A..59H}.}
  \label{fig:bands}
\end{figure*}

Figure \ref{fig:bands} shows the resulting 68\% and 95\% containment regions when combining the results from both sources. The combined intensities from each source's highest likelihood model are shown in black circles along with results from other experiments (\citealt{Abeysekara_2019_VERITAS};\citealt{2019MNRAS.486.4233A}; \citealt{2017A&A...606A..59H}). Everything else in the plot is identical to what is shown in figure \ref{fig:bands_sep}. From figure \ref{fig:bands} it is clear that for some $\lambda$-values, namely $30\mu m<\lambda$ for Mrk\,421 and $\lambda<4 \mu m$ $\cup$ $10\mu m<\lambda$ for Mrk\,501,
boundaries cannot be established (see section \ref{sec:3ML}), limits in these cases are not reported (these are shown as a dash in table \ref{tab:limits}).

The containment bands are in good agreement with the region delimited between the lower and upper limits from galaxy counts and direct measurements respectively. However, for wavelengths lower than $\lambda$=$\sim$5$\mu$m, the containment region is broader than the limits, so results are non-constraining in this range.  
Combined results show a general downward tendency in the EBL measurement for $\lambda$ values $>$5$\mu$m, following the lower limits from galaxy counts. This downward trend has been observed also by VERITAS \citep{Abeysekara_2019_VERITAS} and MAGIC \citep{2019MNRAS.486.4233A}. In the case of HAWC, this could be explained by the fact that the highest observed data points ($>7$ TeV), especially for Mrk\,501, show an upward trend (see \cite{albert2021longterm}), leading to lower EBL density values for this wavelength range. HAWC observed spectrum of Mrk\,421 shows a cutoff energy around 5 TeV, this could, in principle, be related to the bump around 20$\mu$m seen in Mrk\,421 results (see figure \ref{fig:bands_sep}).

\begin{table*}
\caption{Mrk\,421 and Mrk\,501 combined 68\%  and 95\% credible limits for different $\lambda$ values. Note: dash indicates that a limit could not be established.}
\centering
\begin{tabular}{lccccc}
\hline 
$\lambda$  & $\nu$I$_{min}$(68\% CL) & $\nu$I$_{max}$(68\% CL) & $\nu$I$_{min}$(95\% CL) & $\nu$I$_{max}$(95\% CL) \\
$\mu$m  & nW m$^{-2}$ sr$^{-1}$ & nW m$^{-2}$ sr$^{-1}$  & nW m$^{-2}$ sr$^{-1}$ & nW m$^{-2}$ sr$^{-1}$\\
\hline

1.0  &  6.68  &  -  &  4.49  &  -  \\
1.52  &  5.64  &  -  &  4.01  &  -  \\
2.31  &  6.68  &  -  &  4.76  &  -  \\
3.51  &  6.31  &  -  &  4.49  &  -  \\
5.34  &  3.79  &  15.48  &  2.28  &  -  \\
8.11  &  2.28  &  10.76  &  -  &  15.41  \\
12.33  &  1.71  &  9.79  &  -  &  14.02  \\
18.74  &  2.7  &  12.74  &  -  &  19.91  \\
28.48  &  -  &  8.36  &  -  &  11.87  \\
43.29  &  -  &  6.2  &  -  &  8.73  \\
65.79  &  -  &  7.24  &  -  &  10.46  \\
100.0  &  -  &  5.59  &  -  &  8.2  \\

\hline 

\end{tabular}
\label{tab:limits}
\end{table*}



\subsection{Systematic uncertainties}
\label{sec:systematics}
\begin{figure}
     \centering
     \includegraphics[width=0.5\textwidth]{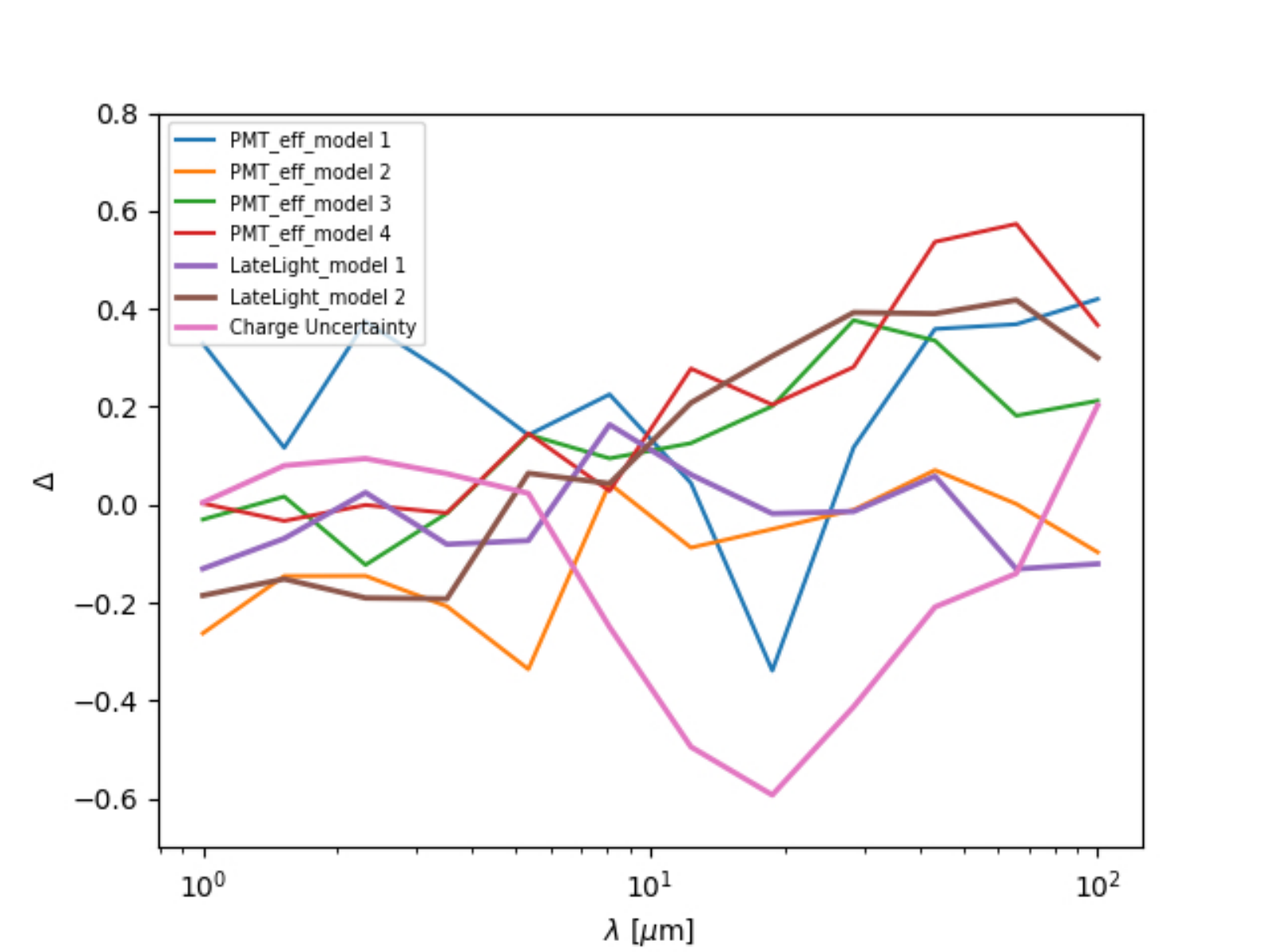}
     \caption{Relative difference between the results considering extreme systematics and the nominal results for different values of wavelength ($\lambda$). The meaning of $\Delta$ is given by equation \ref{eq:diff}. }
     \label{fig:Systematics}
 \end{figure}

There are many sources of systematic uncertainties that affect the flux estimation from HAWC observations, mostly coming from discrepancies between data and simulations, arising from mis-modeling of the detector. A complete treatment of HAWC possible systematics is presented in \citet{Abeysekara_2019}, in which they found that the dominant systematic uncertainties for the spectral flux come from mis-modeling the late light in the air shower and the uncertainty in the modeling of the PMT efficiencies and the charge measured by the PMTs. 
To estimate the potential effect of these systematics in the overall results, the analysis is repeated simulating extreme detector responses considering these dominant systematics. Results in each case are compared with nominal results. Figure \ref{fig:Systematics} shows the relative difference for each wavelength between each considered systematic and the nominal results quantified in the following manner:
\begin{equation}
    \Delta = \frac{\nu I_{\nu}^{nom}-\nu I_{\nu}^{sys}}{\sqrt{\sigma_{nom}^2+\sigma_{sys}^2}}.
    \label{eq:diff}
\end{equation}
where $I_{\nu}^{nom}$ and $I_{\nu}^{sys}$ are the EBL density values, at each wavelength, resulting from the nominal model and the considered extreme systematic model respectively; $\sigma_{nom}^2$ and $\sigma_{sys}^2$ are the corresponding errors, at each wavelength, for the nominal and systematic results respectively.

No significant ($\Delta>$ 3) difference is observed between the results from the nominal model and those coming when considering extreme systematics. It is important to note that the fact systematics are not significant is, in this case, mostly due to the relatively large errors associated with the method, rather than the instrumental uncertainty being small. 


\subsection{Intrinsic model: potential bias}
The selection of a particular emission model, magnetic field and emission radius (see section \ref{sec:IntSpec}) introduces a bias that will eventually impact the calculated EBL limits. As mentioned in section \ref{sec:IntSpec}, the choice on the parameter values and model is based on previous studies performed on the markarians (\citealt{albert2021longterm}; \citealt{Mrk421_SSC}; \citealt{Abdo_2011_501}). To estimate the potential bias coming from the selection of the emission model, the analysis was repeated using an electron distribution following a broken powerlaw (BPL) function instead of the exponentially cut-off powerlaw as the lepton distribution for the emission model. In addition, two models following the contours of the $\pm$1$\sigma$ confidence band (see figure \ref{fig:Int_Mrk421}) given by the fit described in section \ref{sec:IntSpec}, are also considered. These latter models are obtained by sampling the fit parameters of the nominal model within their corresponding $\pm$1$\sigma$ errors and retrieving the minimum and maximum flux values for different values of energy. These flux points are then refitted using \textit{Naima} to get the corresponding parameters of the $\pm 1\sigma$ models to be used in the analysis in an analogous way as performed with the nominal model. 

\begin{figure}
     \centering
     \includegraphics[width=0.5\textwidth]{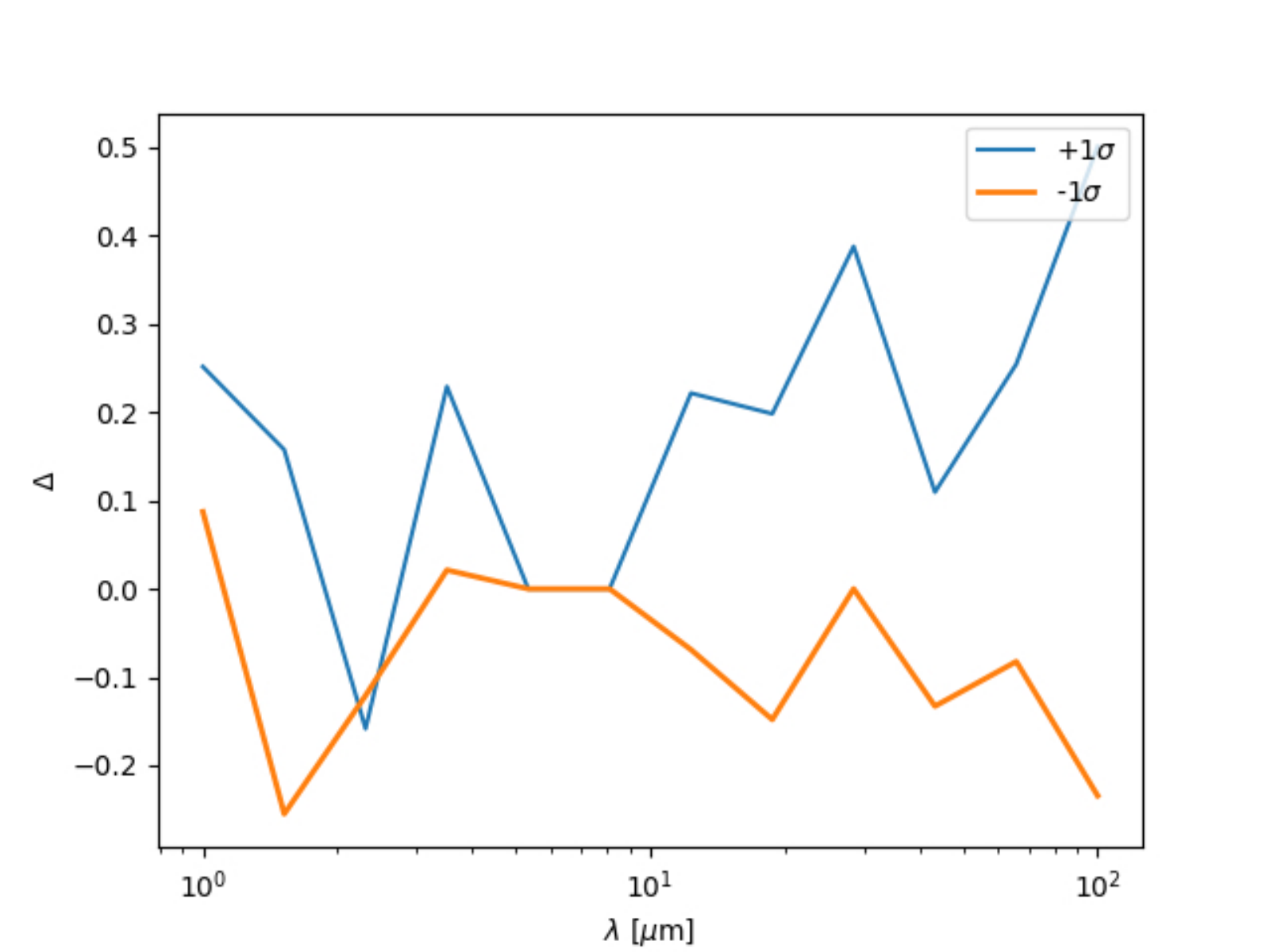}
     \caption{Relative difference between $\pm\sigma$ models, the alternative emission model with a BPL leptonic distribution and the  nominal model, for different wavelengths ($\lambda$). The meaning of $\Delta$ is given by equation \ref{eq:diff}.}
     \label{fig:IntDiff}
 \end{figure}

 The analysis described in section \ref{sec:method} is repeated for these $\pm$1$\sigma$ models and the corresponding results are used to quantify the overall effect of the intrinsic uncertainties in the final containment bands. Figure \ref{fig:IntDiff} shows the relative difference in the derived limits when considering these models with respect to the nominal model and computed as in section \ref{sec:systematics} by using equation \ref{eq:diff}. No significant change is observed between the results of the nominal model and the ones resulting from assuming the $\pm$1$\sigma$ models and assuming the BPL lepton distribution. This suggests that potential biases introduced by the assumptions in the intrinsic emission model don't have a significant impact in the final results within this wavelength range.

 \section{Conclusions}
 After years of operation, the HAWC observatory is able to significantly detect  extragalactic sources, like the Markarians, up to energies of around 10 TeV. This motivates an analysis that could, in principle, probe the mid-IR region of the EBL, a region that is inaccessible to other gamma-ray instruments.
 In this study, physically motivated emission models and \textit{Fermi-LAT} data are used to construct an intrinsic spectrum for each of the Markarians, using the \textit{Naima} python package. A large number of EBL model shapes are randomly generated and used to obtain ``observed" spectra to compare with HAWC data of the Markarians. The comparison is performed using \textit{threeML}, in a way that each EBL model can be assigned a likelihood value that expresses the agreement between that particular spectral realization and HAWC data. The present method has the advantage of being independent of any particular EBL shape and assuming physically motivated emission models as intrinsic spectra. 
 The EBL intensity measurements and containment bands are calculated from 1$\mu$m to 100$\mu$m, probing higher wavelength values than previous measurements performed with VERITAS by using a similar method \citep{Abeysekara_2019_VERITAS}. Results for both sources are in agreement with current upper limits from direct IR observations and lower limits from galaxy counts. Limits are, in general, less constraining for wavelengths below $\lambda$=$\sim$5$\mu$m and there is a downward trend when moving to higher $\lambda$ values, roughly following the lower limits. This trend is clearer in the case of Mrk\,501. A bump around 20$\mu$m is observed for Mrk\,421, possibly due to an observed cutoff drop present in the source observed spectrum around $\sim$5 TeV \cite{albert2021longterm}. Results are also in agreement with other instruments EBL intensity estimations reported by HESS \citep{2017A&A...606A..59H} and MAGIC \citep{2019MNRAS.486.4233A}.
 
 
 These results would surely improve their constraining power if more adequate sources were included. HAWC keeps collecting data from extragalactic sources that could soon be used to derive EBL limits by applying the present method. The radio galaxy M87 was pointed out by \cite{Franceschini_2019_IR_IdeallSource} as a good candidate to perform EBL related studies. At the time of writing, this source was detected at too low significance to be reliable for the analysis, therefore it is not included. However, M87 data are being accumulated \citep{2021_HAWC_Survey} and will soon provide a significant detection that will make the source useful for this type of analysis.

\section*{Acknowledgments}
We acknowledge the support from: the US National Science Foundation (NSF); the US Department of Energy Office of High-Energy Physics; the Laboratory Directed Research and Development (LDRD) program of Los Alamos National Laboratory; Consejo Nacional de Ciencia y Tecnolog\'ia (CONACyT), M\'exico, grants 271051, 232656, 260378, 179588, 254964, 258865, 243290, 132197, A1-S-46288, A1-S-22784, c\'atedras 873, 1563, 341, 323, Red HAWC, M\'exico; DGAPA-UNAM grants IG101320, IN111716-3, IN111419, IA102019, IN110621, IN110521; VIEP-BUAP; PIFI 2012, 2013, PROFOCIE 2014, 2015; the University of Wisconsin Alumni Research Foundation; the Institute of Geophysics, Planetary Physics, and Signatures at Los Alamos National Laboratory; Polish Science Centre grant, DEC-2017/27/B/ST9/02272; Coordinaci\'on de la Investigaci\'on Cient\'ifica de la Universidad Michoacana; Royal Society - Newton Advanced Fellowship 180385; Generalitat Valenciana, grant CIDEGENT/2018/034; Chulalongkorn University’s CUniverse (CUAASC) grant; Coordinaci\'on General Acad\'emica e Innovaci\'on (CGAI-UdeG), PRODEP-SEP UDG-CA-499; Institute of Cosmic Ray Research (ICRR), University of Tokyo, H.F. acknowledges support by NASA under award number 80GSFC21M0002. We also acknowledge the significant contributions over many years of Stefan Westerhoff, Gaurang Yodh and Arnulfo Zepeda Dominguez, all deceased members of the HAWC collaboration. Thanks to Scott Delay, Luciano D\'iaz and Eduardo Murrieta for technical support.

\bibliographystyle{aasjournal}
\bibliography{biblio_fernandez}
\end{document}